%% file: main.tex
\documentclass[conference]{IEEEtran}

\usepackage{url}
\usepackage[hidelinks]{hyperref}

\usepackage{graphicx}
\usepackage[dvipsnames]{xcolor}

\usepackage{algorithm}
\usepackage{algpseudocode}
\usepackage[beginLComment=//~,endLComment=]{algpseudocodex}
\algnewcommand{\LineComment}[2]{\vspace{0.5em} \Statex \hspace{#1em} \textcolor{gray}{\textit{// #2}}}

\usepackage{amsmath}
\usepackage{tikz}
\usepackage{booktabs}
\usepackage{multirow}

\usepackage{listings}
\definecolor{codegreen}{rgb}{0,0.6,0}
\definecolor{codegray}{rgb}{0.5,0.5,0.5}
\definecolor{codepurple}{rgb}{0.58,0,0.82}
\definecolor{backcolour}{rgb}{0.95,0.95,0.92}
\lstdefinestyle{mystyle}{
    backgroundcolor=\color{backcolour},   
    commentstyle=\color{codegreen},
    keywordstyle=\color{magenta},
    numberstyle=\tiny\color{codegray},
    stringstyle=\color{codepurple},
    basicstyle=\ttfamily\footnotesize,
    breakatwhitespace=false,         
    breaklines=true,                 
    captionpos=b,                    
    keepspaces=true,                 
    numbers=left,                    
    numbersep=5pt,                  
    showspaces=false,                
    showstringspaces=false,
    showtabs=false,                  
    tabsize=2
}
\lstdefinelanguage{yaml}{
    basicstyle=\color{BlueViolet}\small,
    rulecolor=\color{black},
    sensitive=false, %
    string=[s]{'}{'},
    stringstyle=\color{BlueViolet},
    morestring=[b]{"}, %
    columns=fullflexible,
    keywords={true,false,null,y,n},
    keywordstyle=\color{PineGreen},
    keywordstyle=[2]\color{BlueViolet}, %
    numbers=none,
    showstringspaces=false,
    breaklines=true,
    frame=top bottom,
    comment=[l]{:},
    morecomment=[s]{/*}{*/},
    commentstyle=\color{RubineRed}
}

\usepackage{caption}
\DeclareCaptionFont{black}{ \color{black} }
\DeclareCaptionFormat{listing}{
  \parbox{\textwidth}{\hspace{15pt}#1:#3}
}

\usepackage{cite}
\usepackage{amsmath,amssymb,amsfonts}
\usepackage{textcomp}
\def\BibTeX{{\rm B\kern-.05em{\sc i\kern-.025em b}\kern-.08em
    T\kern-.1667em\lower.7ex\hbox{E}\kern-.125emX}}

\newcommand\copyrighttext{%
  \footnotesize \textcopyright 2025 IEEE. Personal use of this material is permitted. Permission from IEEE must be obtained for all other uses, in any current or future media, including reprinting/republishing this material for advertising or promotional purposes, creating new collective works, for resale or redistribution to servers or lists, or reuse of any copyrighted component of this work in other works. DOI: \href{https://doi.org/10.1109/LCN65610.2025.11146363}{10.1109/LCN65610.2025.11146363}}
\newcommand\copyrightnotice{%
\begin{tikzpicture}[remember picture,overlay]
\node[anchor=south,yshift=10pt] at (current page.south) {\fbox{\parbox{\dimexpr\textwidth-\fboxsep-\fboxrule\relax}{\copyrighttext}}};
\end{tikzpicture}%
}

\begin{document}

\title{A Bayesian Network Approach for Enhancing Security-Focused Decision Support Systems}

\author{
\IEEEauthorblockN{
Carolina Fern\'{a}ndez-Mart\'{i}nez\IEEEauthorrefmark{1}\IEEEauthorrefmark{2},
Shuaib Siddiqui\IEEEauthorrefmark{1},
Vanesa Daza\IEEEauthorrefmark{2}
}
\IEEEauthorblockA{
    \IEEEauthorrefmark{1}Department of Cybersecurity, i2CAT Foundation, \mbox{08034 Barcelona, Spain}\\
    \{carolina.fernandez, shuaib.siddiqui\}@i2cat.net\\
    \IEEEauthorrefmark{2}Department of Information and Communication Technologies, Universitat Pompeu Fabra, \mbox{08018 Barcelona, Spain}\\
    vanesa.daza@upf.edu\\
}
}

\maketitle
\copyrightnotice

\input{abstract.tex}

\IEEEpeerreviewmaketitle

\input{sections/01-introduction}
\input{sections/02-related_work}
\input{sections/03-background}
\input{sections/04-system_arch}
\input{sections/05-application}
\input{sections/06-validation}
\input{sections/07-conclusions}

\bibliographystyle{IEEEtran}
\bibliography{references}

\end{document}

%% file: abstract.tex
\begin{abstract}
The adoption and integration of heterogeneous stacks in most of today's open-source based networks brings clear benefits like interoperability and availability of advanced features. Yet, on the other hand the increasing number of interconnecting components and moving parts requires maintaining an ever increasing base of interdisciplinary knowledge of different tools in different domains to ensure proper operation.
To alleviate such efforts, this work proposes a Decision Support System (DSS) to guide infrastructure operators through the selection of security approaches (e.g. tools) to adopt in their environments. This framework easily captures the end-user high-level requirements on the security triad for different domains and runs inference on the designated models to provide the identified tools (security mechanisms) that better serve such needs.
The presented DSS aims at delivering an understandable and extensible framework to accommodate varying requirements and Bayesian Network (BN) models.
The architecture and modelling of the system are proposed, aligned with its theoretical framework.
Its performance is evaluated in terms of time and prediction accuracy.
\end{abstract}


%% file: sections/01-introduction.tex
\section{Introduction}

Today’s networks and IT environments exhibit unprecedented densification, driven by the increased adoption of virtualisation and softwarisation (e.g. microservices) architectures and cloud-native paradigms that permeate cloud and telecommunication environments.
These environments integrate multiple virtualised, complex stacks to serve core functionalities, such as identity management.

As organisations expand and combine these approaches, their infrastructure becomes an amalgam of different components.
This is particularly evident in setups integrating self-hosted stacks (i.e. private cloud), and also between local and remote service providers (i.e. public cloud) environments. It becomes also increasingly relevant, given recent intention to return workloads from public to private clouds \cite{Lenschow_Kalia_Sandler_El-Assal_2024}.
Naturally, this introduces operational complexity due to varying configuration and service delivery approaches per stack, even if partially mitigated by separating concerns.
The general knowledge gap for system operators is widened, aggravated by the lack of qualified cybersecurity experts \cite{Ahmed_Hossain_Fazio_Lezzi_Islam_2024}.
Additionally, (security) misconfiguration is increasing, enlarging environments' vulnerability surface \cite{LOUREIRO202113}; which often rely on several interconnected open interfaces.
To ensure interdisciplinary, efficient and safe operations; organisations can either seek external support or invest internally to acquire expertise\cite{10.1145/3243734.3243794}.

To foster internal investment, Decision Support Systems (DSS) can initially guide the operator's actions, focusing on investigating and shortening the search for specific logic.
This poses multiple challenges, e.g. on eliciting security requirements in a manner that balances simplicity and understandability for the operator and generalising the model definition so to cover other logical domains or non-security dimensions (e.g. regulatory or financial requirements).

The proposed work aims to bridge the operator's security knowledge gap by suggesting ranked Security Mechanisms for adoption, reducing time and complexity.
This work contributes towards that objective by proposing (i) a simple and extensible end-user requirement syntax, pondering each Security Dimension from the CIA (Confidentiality, Integrity, Availability) triad;
(ii) a syntax to model the discrete Bayesian Network (BN) \cite{Jensen_Nielsen_2007} in a more human-understandable and extensible way; and
(iii) a framework to interpret and load required data per Security Domain through Knowledge Graphs, implemented as discrete multi-level BNs that permit statistical inference to propose the most Suitable Security Mechanisms.

The paper is structured as follows: Section \ref{sec:related-work} discusses the related State of the Art; Section \ref{sec:background} introduces the contribution's theoretical framework, which is detailed in Section \ref{sec:system-arch}. Section \ref{sec:application} applies it to practice; Section \ref{sec:validation} evaluates it; and Section \ref{sec:conclusions} summarises the contribution and next steps.

%% file: sections/02-related_work.tex
\section{Related work}
\label{sec:related-work}

DSS systems are applied in virtually every discipline, from medicine to engineering, including cybersecurity.
Surveyed approaches are grouped by technique in Table \ref{tab:related-work:comparison}.

Following Dynamic Programming (DP), \cite{OZDEMIRSONMEZ2022102865} and \cite{SCHMIDT2021107093} perform Multi-Objective Optimisation (MOO), while seeking an acceptable trade-off. \cite{OZDEMIRSONMEZ2022102865} uses Mixed Integer Linear Programming (MILP) to find efficient security controls (or mechanisms) for healthcare environments; mapping these to vulnerabilities, creating an Attack Graph and assigning relative costs (constraints) to each.
\cite{SCHMIDT2021107093} uses Integer Linear Programming (ILP) to minimise cost and risk and selects security controls based on organisational risk preference.
In \cite{PAUL2021349}, Non-Linear Programming (NLP) minimises a multi-constrained objective function, reducing cybersecurity investment.
For Industry 4.0, \cite{Sawik_2022} applies NLP and transforms it into MILP to determine security controls, considering multiple constraints to minimise economic investment and losses.

Graph-based approaches model events and their relationships for further analysis through graph traversal and interpretation. General service graphs can model inter-dependencies and impacts across services for security management \cite{8842769}; or cybersecurity challenges and influences as digraphs \cite{Ahmed_Hossain_Fazio_Lezzi_Islam_2024}, incorporating combinatorial and matrix algebra for prioritisation.

\begin{table}[t]
\vspace{0.4em}
\caption{Literature review according to adopted technique}
\centering
\begin{tabular}{lc}
    \textbf{Technique type} & \textbf{Contributions} \\[0.3em]
    \toprule
    Multi-Objective Optimisation (MOO) & \cite{OZDEMIRSONMEZ2022102865} (MIP) \cite{SCHMIDT2021107093} (IP) \\
    Dynamic Programming & \cite{PAUL2021349} (NLP) \\
    Probabilistic Graphical Models & \cite{Wang_Neil_Fenton_2020} (BN) \cite{10.1007/978-3-642-16644-0_39} (BN) \\ 
    & \cite{Hunte_Neil_Fenton_2024} (HBN) \cite{Simsek_Dag_Coussement_Kibis_Asilkalkan_Ragothaman_2025} (TAN) \\
    Stochastic & \cite{doi:10.1177/1548512916683451} (MC) \cite{Zadeh_Jeyaraj_2022} (MC) \\
    \bottomrule
\end{tabular}
\label{tab:related-work:comparison}
\end{table}

Formal approaches, such as Markov Chains (MC), can assist cybersecurity analysts by encoding stochastic processes and their transitions. For instance, to predict transitions from observed to unknown cyber threats \cite{doi:10.1177/1548512916683451}; or use reported risk data and scores to assess the organisation's security posture and balance exploration and exploitation capabilities \cite{Zadeh_Jeyaraj_2022}.
In particular to cybersecurity-focused, BN-based DSS systems, \cite{10.1007/978-3-642-16644-0_39} evaluates the severity of the detected threats, informing critical infrastructure operators to minimise risks and simulate potential actions.
BNs are used for rigorous modelling to complement well-known risk assessment algorithms to circumvent inherent approximations for e.g. medical \cite{Wang_Neil_Fenton_2020} and consumer \cite{Hunte_Neil_Fenton_2022} devices' risk assessment.
Hybrid Bayesian Network (HBN) modelling combines BNs and Decision Trees to detect potential hazards in medical devices \cite{Hunte_Neil_Fenton_2024}; offering dynamic inference and higher accuracy compared to simpler BNs.
Tree-Augmented Networks (TAN) are used in e.g. financial fraud detection \cite{Simsek_Dag_Coussement_Kibis_Asilkalkan_Ragothaman_2025} to encode conditional-dependency relationships and integrate Elastic net and Random Forest-based genetic algorithms for hybrid classification.

The proposed work applies BNs before risk assessment or threat detection stages, delivering a multi-layer BN-based solution that uses statistical inference to compute the most suitable security decision.
This enables a simple, versatile definition of both the security end-user requirements (based on the CIA triad) and the knowledge base (modelling the impact between Security Mechanisms, Categories and Requirements).

%% file: sections/03-background.tex
\section{Background}
\label{sec:background}

The proposed DSS relies on BNs, a well-known Probabilistic Graphical Model (PGM) that applies Bayes' theorem and Bayesian inference to calculate probabilities for a subset of events.
A BN can represent and compute the probabilities of any event or random variable $X_i \in (X_1, ..., X_n)$, conditionally (in)dependent of each other.
A BN contains (i) a Directed Acyclic Graph (DAG), encoding vertices (nodes) from $X_i$ and directed edges from the influence of random variables $X_i$ to $X_j$; and (ii) Conditional Probability Distributions (CPD) with the probability of $X_i$ occurring conditioned on that of its predecessors (parent) nodes. That is, $BN = DAG \times CPD$.

Through Bayesian statistical inference, a BN can compute (infer) the probability of an event or observation (node in the DAG), conditioned on others' probabilities.
On the one hand, Bayes' theorem (\ref{eq:background:bayes-theorem}) computes the hypothesis (H) probability from the probabilities on already known data, or evidence (E).

\begin{equation}
\label{eq:background:bayes-theorem}
P(H|E) = \frac{P(E|H) \cdot P(H)}{P(E)}
\end{equation}

On the other hand, to apply the Bayes' theorem (\ref{eq:background:bayes-theorem}), the DAG is traversed to grasp the influence between each pair of involved nodes $(X_i, X_j)$. Subsequently, the chain rule (\ref{eq:background:bayes-chain-rule}) is used to simplify calculations (e.g. \emph{d-separation}) and to compute the final result from the joint probability of random variables.
In each step, it computes the probability of a node $X_i$ conditioned on the probabilities of its parent nodes $pa(X_i)$.

\begin{flalign}
\label{eq:background:bayes-chain-rule}
\begin{aligned}
&P(X_1, ..., X_n) = P(X_1 \cap ... \cap X_n) = \prod_{i=1}^n {P(X_i ~|~ pa(X_i))}
\end{aligned}
\end{flalign}

Finally, BNs can use different reasoning patterns \cite{Koller_Friedman_2010}, such as (i) predictive (or causal), forecasting the likelihood of an effect to occur given a cause; and (ii) diagnosis (or evidential), determining the chances of an already occurred effect due to a given cause \cite{Requejo_Castro_Perez_2021}.
Besides these, (iii) intercausal (or \textit{explaining away}) identifies the contributions to a given effect by multiple competing causes. Some \cite{Reasoning_with_Bayesian_Belief_Networks_2016} also consider (iv) decision-making by selecting actions based on the effects' probabilities and on BNs' decision and utility nodes.
Processing varies according to the reasoning pattern: queries interpreting data (evidence for effects, hypotheses for causes) and traversing the BN's DAG differently \cite{L_Crowley_Reasoning_BN_2018}: \textit{top-down} for prediction (cause to effect) or \textit{bottom-up} for diagnosis (effect to cause).

Thus, BNs are suitable due to their representation of the contribution or impact across random variables or events, their non-cyclic nature and their different reasoning patterns.
Here, random variables can be Security Requirements, Categories or Mechanisms; all having a certain impact on others.

%% file: sections/04-system_arch.tex
\section{Proposed system}
\label{sec:system-arch}

The proposed system features a simple syntax to (i) gather end-user Security Requirements and configuration (Section \ref{sec:system-arch:system-arch:reqs-cfg}); (ii) represent multiple BN layers (Section \ref{sec:system-arch:system-arch:bn-models}); and a (iii) DSS to apply Bayesian predictive \cite{Requejo_Castro_Perez_2021} inference on such BNs to extract the most likely (i.e. best fitting) Security Categories and Mechanisms to meet end-user requirements (Section \ref{sec:system-arch:system-arch:algorithm}).
Table \ref{tab:system-arch:nomenclature} explains the system's notation.

For the DSS to operate with multiple considerations (from network security to backup and redundancy), eight Security Domains $D$ were identified. For each, a number of Security Categories $C$ and Security Mechanisms $M$ were identified, based on expert knowledge.
To apply the DSS to a practical domain, two Security Domains were selected: \textit{Authentication and Identity (AI)} and \textit{Network Security (NS)}. Each Security Domain has related Security Categories, such as \textit{Access Control (AC)}, \textit{Segment Configuration (SC)} or \textit{Network Virtualisation (NV)}. Also, each Security Category has related Security Mechanisms, such as \textit{Identity and Access Management (IAM)} or \textit{Firewall (FW)}.

Fig. \ref{fig:system-arch:dss-arch} contains all the Security-focused DSS modules, ordered by step. First,  a \raisebox{.5pt{\textcircled{\raisebox{-.9pt} {1}}} \textit{Parser} module interprets} the models' definitions and the \raisebox{.5pt{\textcircled{\raisebox{-.9pt} {2}}}} ~configuration $F$ and end-user requirements $R$ (i.e. weights on the CIA triad); then performs \raisebox{.5pt}{\textcircled{\raisebox{-.9pt} {3}}} requirement's weight post-processing (Section \ref{sec:system-arch:system-arch:reqs-cfg}).
The \raisebox{.5pt{\textcircled{\raisebox{-.9pt} {4}}}} ~\textit{Model constructor} loads a BN model per security domain $D$ (Section \ref{sec:system-arch:system-arch:bn-models}).
With these, the \textit{Suggester} \raisebox{.5pt}{\textcircled{\raisebox{-.9pt} {5}}} adapts and \raisebox{.5pt}{\textcircled{\raisebox{-.9pt} {6}}} infers the most suitable Security Mechanisms $M$ per end-user Security Requirement and Domain. These outcomes are finally \raisebox{.5pt{\textcircled{\raisebox{-.9pt} {7}}} refined/filtered based on $F_p$}.
Points \raisebox{.5pt{\textcircled{\raisebox{-.9pt} {3}}}-\raisebox{.5pt}{\textcircled{\raisebox{-.9pt} {7}}} are covered in Alg. \ref{alg:system-arch:algorithm:dss}, lines \ref{alg:system-arch:algorithm:dss:line:s10}-\ref{alg:system-arch:algorithm:dss:line:s23}.}

\begin{table}[t]
  \vspace{0.4em}
  \caption{Nomenclature for the proposed DSS system}
  \centering
  \begin{tabular}{cl}
    \toprule
    \textbf{Term} & \textbf{Description} \\
    \hline \\[-0.7em]
    \multicolumn{2}{l}{\textbf{Inputs}} \\[0.3em]
    $D$, $d$ & Security Domains. Ex.: \emph{ai}, \emph{br}, \emph{ns} \\
    $F$ & Configuration. $F_p \in (0,1)$ limits no. results ($M$) \\
    $R$, $r$ & Security Requirements. Ex.: \emph{(C,I,A)} $\in (0,1)^3$ \\
    $C$, $c$ & Security Categories. Ex.: \emph{ac}, \emph{authn}, \emph{sc} \\[0.5em]
    \multicolumn{2}{l}{\textbf{Output} (per domain $\in D$)} \\[0.5em]
    $M_{d}$ & Security Mechanisms. Ex.: \emph{iam}, \emph{fw} \\[0.5em]
    \multicolumn{2}{l}{\textbf{Internals} (per layer $X$ and/or domain $\in D$)} \\[0.5em]
    $BN_{lX,d}$ & BN object\\
    $RV_{lX,d}$ & Random variables of a BN model\\
    $VP_{d}$ & Cartesian product of random variables\\
    \bottomrule
  \end{tabular}
  \label{tab:system-arch:nomenclature}
\end{table}

\subsection{Requirements and configuration definition}
\label{sec:system-arch:system-arch:reqs-cfg}

End-user requirements are described in simple, concise YAML files (Listing \ref{lst:system-arch:cfg:requirements}).
The operator indicates the importance of each security dimension, i.e. every element in the CIA triad; resulting in a triplet $R$ for each Security Domain ($D$).

\begin{figure}[!h]
\centering
\begin{minipage}{0.40\columnwidth}
\begin{lstlisting}[label=lst:system-arch:cfg:requirements,caption=Weighed and ranked requirements,language=yaml]
requirements:
  domain:
    ai:
      confidentiality: 0.3
      integrity: 0.5
      availability: 0.2
    ns:
      confidentiality: 1
      integrity: 0
      availability: 2
\end{lstlisting}
\end{minipage}
\hfill
\begin{minipage}{0.56\columnwidth}
\begin{lstlisting}[label=lst:system-arch:cfg:general,caption=General configuration,language=yaml]
general:
  ...
  results:
    percentile: 0.45
models:
  categories:
    spec: "models/categories.yaml"
  mechanisms:
    spec: "models/mechanisms.yaml"
\end{lstlisting}
\end{minipage}
\end{figure}

\begin{figure}[t]
    \centering
    \includegraphics{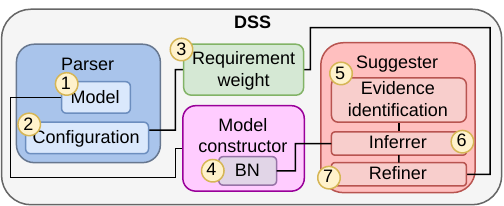}
    \caption{DSS logical architecture}
    \label{fig:system-arch:dss-arch}
\end{figure}

Security Requirements ($R$) are provided to the \textit{Parser module (Fig. \ref{fig:system-arch:dss-arch}, steps 1-2) in} two ways:
(i) weighted, as real numbers between 0 and 1 whose sum equals 1, e.g. $(0.3, 0.5, 0.2)$; and (ii) ranked, starting in 1, as an increasing monotonic sequence, e.g. $(2, 1, 3)$.
Any (but not all) dimensions can be set to zero.
Each ranked value ($r_d$) is internally normalised to the weighted ($r_{d,w}$) approach, applying \ref{eq:system-arch:algorithm:weight-ranking} in Alg. \ref{alg:system-arch:algorithm:dss}, line \ref{alg:system-arch:algorithm:dss:line:s11} within the \textit{Requirement weight module} (Fig. \ref{fig:system-arch:dss-arch}, step 3).

\begin{equation}
\label{eq:system-arch:algorithm:weight-ranking}
r_{d,w} = \frac{|R| - r_{d} + 1}{\sum_{d} r_{d}} ~ \forall ~ r_{d} \in R ~,~ d \in D
\end{equation}

The user can also provide the result percentile $F_p$, which acts as a filtering threshold (Listing \ref{lst:system-arch:cfg:general}).
The results output by the algorithm are first sorted, then filtered from the best (i.e. first) until the $n$-th position; with $n = \lfloor F_{p} \rfloor ~ \cdot ~ |M|$.

\subsection{BN modelling}
\label{sec:system-arch:system-arch:bn-models}

The \textit{Model constructor} (Fig. \ref{fig:system-arch:dss-arch}, step 4) loads BN models from the Configuration ($F$), represented via text-based definitions (Listing \ref{lst:system-arch:model:categories}) of Security (i) Domains, (ii) Categories and (iii) Mechanisms. Some models are pre-bundled, yet can be manually modified before being loaded in the DSS (in Alg. \ref{alg:system-arch:algorithm:dss}, lines \ref{alg:system-arch:algorithm:dss:line:s12} and \ref{alg:system-arch:algorithm:dss:line:s21}).
A BN model is defined by: (i) random variables (here, the DAG's leaves), encoding Security Categories or Mechanisms (depending on each model or layer); along with (ii) links and (iii) CPD tables, expressing impact and quantitative contribution between variables, respectively.
Invalid or missing data (e.g. missing CPDs for joint probability distributions) are notified to the user.
Listing \ref{lst:system-arch:model:categories} and Fig. \ref{fig:system-arch:dss-2layers-bn} show the BN for the \emph{ns} Security Domain.

\begin{figure*}[!bht]
\vspace{0.4em}
\centering
\begin{minipage}{0.32\textwidth}
\begin{lstlisting}[language=yaml,firstline=1,lastline=11]
model:
  ...
    bayesian-network:
      global:
        random-variables:
          g_a: "Availability"
          ...
        bn-type: "DiscreteBayesianNetwork"
        inference:
          variable-elimination: True
          belief-propagation: False
\end{lstlisting}
\hfill
\vspace{0.95em}
\end{minipage}%
\hfill
\begin{minipage}{0.32\textwidth}
\begin{lstlisting}[label=lst:system-arch:model:categories,caption=Categories' BN definition,language=yaml,firstline=12,lastline=22]
      domain:
        ...
        ns:
          random-variables:
            ac:
              name: "Access control"
            ...
          links:
            ...
            - src: sc
              dst: [g_i, g_a]
\end{lstlisting}
\end{minipage}
\hfill
\begin{minipage}{0.32\textwidth}
\begin{lstlisting}[language=yaml,firstline=23,lastline=33]
          cpds:
            ...
            nv:
            ...
            - conditions:
                ac: 0
                sc: 1
              values:
                false: 1/3
                true: 2/3
            ...
\end{lstlisting}
\hfill
\vspace{0.95em}
\end{minipage}
\end{figure*}

\subsection{Inference algorithm}
\label{sec:system-arch:system-arch:algorithm}

\begin{figure}[t]
    \centering
    \includegraphics{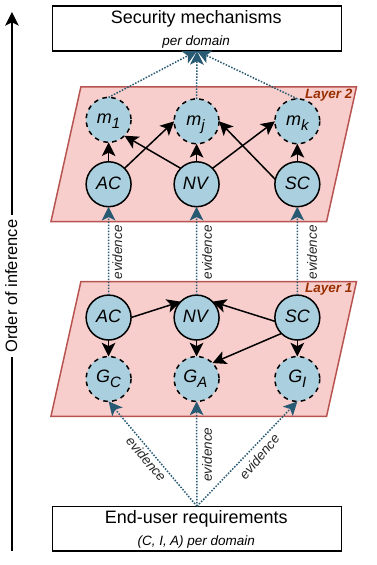}
    \caption{Layers of BNs, part of the BN core}
    \label{fig:system-arch:dss-2layers-bn}
\end{figure}

After running the logic from Section \ref{sec:system-arch:system-arch:reqs-cfg} (Fig. \ref{fig:system-arch:dss-arch}, steps 1-2) and Section \ref{sec:system-arch:system-arch:bn-models} (Fig. \ref{fig:system-arch:dss-arch}, step 4); Alg. \ref{alg:system-arch:algorithm:dss} runs the core inference logic in lines \ref{alg:system-arch:algorithm:dss:line:s13}-\ref{alg:system-arch:algorithm:dss:line:s23} (Fig. \ref{fig:system-arch:dss-arch}, steps 3-7).
Fig. \ref{fig:system-arch:dss-2layers-bn} pictures its two stages. Bottom-up (i.e. Layer 1 to 2), it processes Security Requirements and Security Mechanisms as respective inputs.
Layer 1 maps (i) Security Requirements ($R$) to (ii) Security Categories ($C$); and Layer 2 maps (i) Security Categories (ii) Security Mechanisms ($M$). Each layer returns the best fitting (most likely) output. Layer 2 is fed the output of Layer 1.

\begin{algorithm}
\caption{Find best fitting Security Mechanisms}
\begin{algorithmic}[1]
\Require $D \neq \emptyset; R_{d} \subseteq (C,I,A) \neq \emptyset, 1 \geq r_{d} \geq 0, r \in R, d \in D; F \neq \emptyset; F_{p} > 0$
\Ensure $M_{d} \neq \emptyset, d \in D$

\For{each $d$ in $D$} \label{alg:system-arch:algorithm:dss:line:s10}

    \LineComment{1.3}{Layer 1: find best fitting security categories}
    \State $R_{d,w} \gets normalise\_requirements\_weights(R_{d})$ \label{alg:system-arch:algorithm:dss:line:s11}
    \State $BN_{l1,d} \gets bayesian\_networks(F)$ \label{alg:system-arch:algorithm:dss:line:s12}
    \State $RV_{l1,d} \gets random\_variables(BN_{l1,d})$ \label{alg:system-arch:algorithm:dss:line:s13}
    \State $VP_{d} \gets cartesian\_product(RV_{l1,d})$ \label{alg:system-arch:algorithm:dss:line:s14}
    \For{each $r_{d,w}$ in $R_{d,w}$}
        \State $r_{d,c} \gets quantise\_requirements\_weights(r_{d,w})$ \label{alg:system-arch:algorithm:dss:line:s15}
        \State $C_{l1,d,r} \gets conditional\_probability(r_{d,c} ~|~ VP_{d})$ \label{alg:system-arch:algorithm:dss:line:s16}
        \State $C_{l1,d,r,s} \gets score\_results(C_{l1,d,r}, r_{d,w}, BN_{l1,d})$ \label{alg:system-arch:algorithm:dss:line:s17}
    \EndFor
    \State $C_{l1,d} \gets filtered\_refined\_results(C_{l1,d}, F_{p})$ \label{alg:system-arch:algorithm:dss:line:s18}

    \LineComment{0}{Layer 2: find best fitting security mechanisms}
    \State $BN_{l2,d} \gets bayesian\_network(F)$ \label{alg:system-arch:algorithm:dss:line:s21}
    \For{each $c$ in $C_{l1,d}$}
        \State $C_{l2,d,r} \gets conditional\_probability(c ~|~ C_{l1,d})$ \label{alg:system-arch:algorithm:dss:line:s22}
    \EndFor
    \State $M_{d} \gets filtered\_refined\_results(C_{l2,d,r}, F_{p})$ \label{alg:system-arch:algorithm:dss:line:s23}
\EndFor
\end{algorithmic}
\label{alg:system-arch:algorithm:dss}
\end{algorithm}

In Layer 1 (i.e. Security Categories' model), and as part of the \textit{Evidence identification module} (Fig. \ref{fig:system-arch:dss-arch}, step 5), its BN identifies all possible values per random variable used as evidence (Alg. \ref{alg:system-arch:algorithm:dss}, line \ref{alg:system-arch:algorithm:dss:line:s13}).
The Cartesian product of each value is then computed (Alg. \ref{alg:system-arch:algorithm:dss}, line \ref{alg:system-arch:algorithm:dss:line:s14}); for instance, for 3 random variables with binary values, this yields: $(0, 0, 0), ..., (1, 1, 1)$.
Security Requirements have its weight quantised (Alg. \ref{alg:system-arch:algorithm:dss}, line \ref{alg:system-arch:algorithm:dss:line:s15}); applying $f:\mathbb{R}\rightarrow\mathbb{B}$.

Inference requests (\ref{eq:system-arch:algorithm1:conditioned-prob}) are prepared with (i) each normalised weight for Security Requirements ($r_{d,c}$), as hypothesis; and (ii) permuted values (Cartesian product) for Security Categories ($VP_d = C_j ~ \times ~ C_k$), as evidence.
Such requests run in the \textit{Inferrer module} (Fig. \ref{fig:system-arch:dss-arch}, step 6) in Alg. \ref{alg:system-arch:algorithm:dss}, line \ref{alg:system-arch:algorithm:dss:line:s16}.

\begin{equation}
\label{eq:system-arch:algorithm1:conditioned-prob}
C_{l1,d,r} = P(r_{d,c} ~ | ~ C_j ~ \times ~ C_k) ~ \forall ~ C_{j},C_{k} \in C ~,~ d \in D
\end{equation}

The most likely Security Categories are filtered in the \textit{Refiner module} (Fig. \ref{fig:system-arch:dss-arch}, step 7) in Alg. \ref{alg:system-arch:algorithm:dss}, line \ref{alg:system-arch:algorithm:dss:line:s18}.
It is worth noting that each layer uses different types of values for hypothesis and evidence, depending on its order of inference.

In Layer 2 (i.e. Security Mechanisms' model), inference requests are prepared based on Layer's 1 outcome: (i) each available Security Mechanism ($c$), as hypothesis; and (ii) one or more Security Categories along with their values ($C_d$), as evidence; the latter being the output of the first BN layer (Alg. \ref{alg:system-arch:algorithm:dss}, line \ref{alg:system-arch:algorithm:dss:line:s22}).
The inference process combines the Bayes theorem (\ref{eq:background:bayes-theorem}) and chain rule (\ref{eq:background:bayes-chain-rule}) to compute the likelihood of its associated hypothesis.
With the inferred results, all hypothesis are enriched with a score (Alg. \ref{alg:system-arch:algorithm:dss}, line \ref{alg:system-arch:algorithm:dss:line:s17}). This applies custom logic to weight the given evidence (\ref{eq:system-arch:algorithm1:hypothesis-scoring}) with respect to (i) the given Security Requirements and, depending on the inference method, also with (ii) the BN structure for the involved Security Category.

\begin{equation}
\label{eq:system-arch:algorithm1:hypothesis-scoring}
C_{l1,d,r,s} = \sum_{d} r_{d,w} \cdot C_{l1,d,r} ~ \forall ~,~ d \in D
\end{equation}

Finally, the \textit{Suggester module uses the Configuration ($F_p$) percentile} ($x \in [0,1]$) to filter results (Alg. \ref{alg:system-arch:algorithm:dss}, line \ref{alg:system-arch:algorithm:dss:line:s23}). E.g. $F_p = 0.35$ returns the highest (most probable) 35\% results.

\subsection{Implementation and deployment}
\label{sec:system-arch:implementation}

The DSS is written in Python 3.10.
YAML models are used to easily define Security Requirements (Listing \ref{lst:system-arch:cfg:requirements}), Configuration (Listing \ref{lst:system-arch:cfg:general}) and the BN models (Listing \ref{lst:system-arch:model:categories}).
This helps users to quickly construct on-the-fly BNs by hiding the complexity of defining the CPDs in the underlying \texttt{pgmpy} library \cite{Ankan2015} (version 1.0.0), and allows straightforward integration with graphical clients through commonly used data structures.

Models run the \textit{DiscreteBayesianNetwork class.}
Inference uses the default \textit{VariableElimination} algorithm, given its reduced time compared to the
\textit{BeliefPropagation} one in this scenario.
The proposed system does not require GPU to run, yet the underlying PyTorch library will use it if available.

%% file: sections/05-application.tex
\section{Application}
\label{sec:application}

The proposed DSS is devised as a stand-alone service.
From the received Security Requirements and BN models, it returns a filtered, prioritised set of Security Mechanisms.

\begin{figure}[b]
    \centering
    \includegraphics{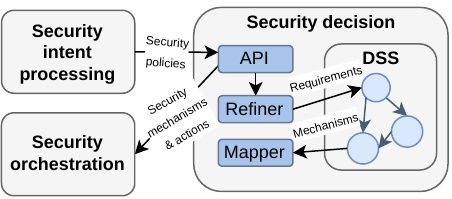}
    \caption{DSS complementing the security processing logic}
    \label{fig:application:6gbricks-arch}
\end{figure}

The DSS is currently under integration with a Zero-Trust framework that setups on-the-fly, secure connections between experimenters and involved experimental infrastructures.
The ZT framework (Fig. \ref{fig:application:6gbricks-arch}) first (i) extracts high-level end-user Security Requirements (i.e. intents); (ii) refines, ponders criteria and maps (through the DSS) the best Security Mechanisms; and (iii) orchestrates the deployment.
In the ZT framework, default security intents explicitly encode Security Categories and Mechanisms. As part of the integration, two methods translate to the internal Security Requirements $R$ triplet: (i) a higher-level form of intents using the same Terse Resource Description Framework (RDF) Triple Language (TTL) standard form, directly mapping to $R$; and (ii) the RDF graph derived from default intents and translated to $R$.

%% file: sections/06-validation.tex
\section{Validation}
\label{sec:validation}

The DSS is evaluated in terms of performance per layer's processing, and on the classification accuracy for Layer 1.

\subsection{Methodology}

Performance tests measure the processing time per group of operations and for each of the two BN layers, to pinpoint those with higher contributions.

Accuracy tests measure the degree of successful classification in Layer 1.
As the initial data (CPD tables) are too small to train a network with, it is synthetically augmented by (i) interpolating closer vectors with the most similar pre-existing CPD entries - obtained through minimum Hamming distance and expanding the discrete ranges for the first BN training (fitting); and (ii) simulating, based on the previously fit BN, to achieve a sensible amount (i.e. 1e3 entries) that can be used for the testing (predicting) stage.
This uses cross-validation with 5 \textit{k-folds} using the default 80\%/20\% training/test ratio. It is then run for both Bayesian and Maximum Likelihood (MLE) estimators.
Different validation scenarios are tested, including combinations of expert-based CPD and partly random data for training and/or testing.

In both cases, a hundred inference queries are run against the DSS and
data is shown aggregated both per Security Domain and layer.
Two Security Domains were selected: \emph{ai} (authentication \& identity) and \emph{ns} (network segmentation).
The two models have comparable sizes, with 3 vertices (Security Categories) and 5 edges (contributions to each other); and the complexity of the CPDs is similar too, where evidence uses 1 or 2 variables.
However, the Security Requirements per security domain vary: 3 were selected for \emph{ai} and 2 for \emph{ns}. 
Manual validation used Netica\footnote{https://www.norsys.com/netica.html} to reconstruct the BN for Layer 1 and domain \textit{ns}. Automated validation uses NumPy and Pandas to process the DSS results, scikit-learn for scoring metrics and classification reports, and Matplotlib for plots.

\subsection{Results and analysis}

Both diagnosis and predictive reasoning were first evaluated per layer and Security Domain \textit{d} (\textit{ai}, \textit{ns}).
Predictive reasoning was chosen, as it takes less time and provides more consistent results.
This is shown in Table \ref{tab:validation:performance:reasoning:prediction:layers}; where average Standard Deviation (SD) for Layer 1 is of 0.00482 and 0.01099 seconds for prediction and diagnosis, respectively (56\% less); and the Inter-Quartile Range (IQR) is of 0.00759 and 0.01293 seconds for prediction and diagnosis, respectively (41\% less).

\begin{figure}[b]
    \centering
    \includegraphics[width=0.5\textwidth]{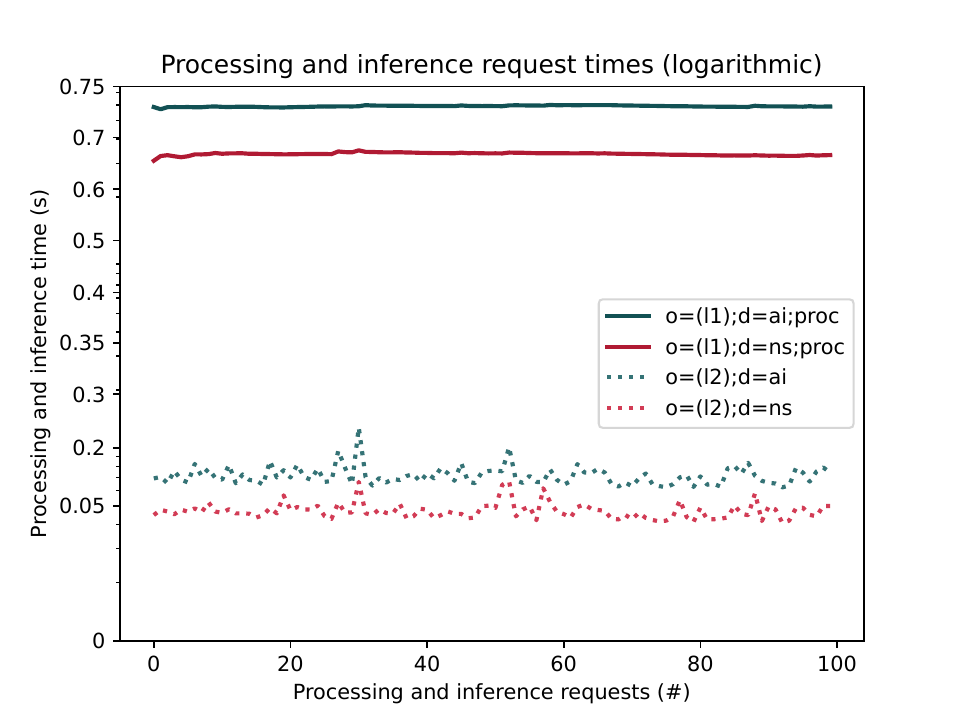}
    \caption{Time (seconds) per layer and (predictive) reasoning}
    \label{fig:validation:reasoning:prediction:layers}
\end{figure}

Fig. \ref{fig:validation:reasoning:prediction:layers} shows the processing times per layer for the predictive reasoning.
Operation times (Alg. \ref{alg:system-arch:algorithm:dss}, lines \ref{alg:system-arch:algorithm:dss:line:s11}-\ref{alg:system-arch:algorithm:dss:line:s18}) for Layer 1 show at the top, and Layer 2 (Alg. \ref{alg:system-arch:algorithm:dss}, lines \ref{alg:system-arch:algorithm:dss:line:s21}-\ref{alg:system-arch:algorithm:dss:line:s23}) at the bottom.
In particular, the Cartesian product (Alg. \ref{alg:system-arch:algorithm:dss}, line \ref{alg:system-arch:algorithm:dss:line:s14}) computes the selected Security Categories (i.e. 3 and 2 for the \emph{ai} and \emph{ns} domains, respectively) and their values (here, 2 per category) for each layer.
E.g. it first finds the 8 possibilities (Section \ref{sec:system-arch:system-arch:algorithm}) and filters out uninteresting combinations like $(0, 0, 0)$; leaving 7.
Thus, the inference process at Layer 1 runs $(2 \cdot 3) \cdot 7 ~=~ 42$ queries to the \emph{ns} domain.
The simplicity of the operations at Layer 2 makes its times negligible compared to those of Layer 1.
The time taken per domain varies per layer, where the one for the \emph{ai} domain is consistently higher than for \emph{ns}. This is expected, as \emph{ns} has Security Requirements associated to 2 Security Categories; while \emph{ai} has 3.
Table \ref{tab:validation:performance:reasoning:prediction:layers} shows data distribution (IQR) and dispersion (SD) are specially accused -with a higher order of magnitude- in Layer 1 and depending on the domain. This is so due to the high number of operations and to the varying structure and/or Security Requirements of each graph.

\begin{figure}[b]
    \centering
    \includegraphics[width=0.5\textwidth]{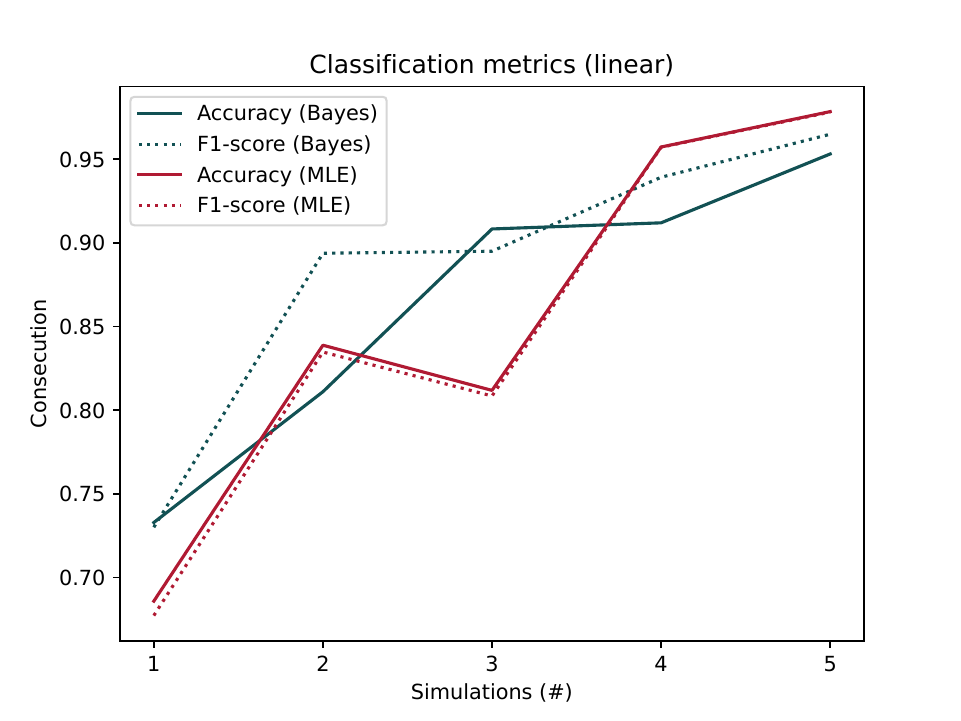}
    \caption{Prediction performance for Layer 1, domain: \textit{ns}}
    \label{fig:validation:accuracy:reasoning:prediction:layers}
\end{figure}

\begin{table}[t]
  \vspace{0.4em}
  \caption{Processing and inference times (seconds)}
  \centering
  \begin{tabular}{lcccc}
    \toprule
    \textbf{BN layer} & \multicolumn{2}{c}{\textbf{1 (Sec. Categories, $C$)}} & \multicolumn{2}{c}{\textbf{2 (Sec. Mechanisms, $M$)}} \\[0em]
    \textbf{Sec. Domain, $D$} & \textbf{ai} & \textbf{ns} & \textbf{ai} & \textbf{ns} \\
    \hline \\[-0.5em]
    \textbf{Minimum} & 0.531 & 0.416 & 0.006 & 0.004 \\
    \textbf{Mode} & 0.531 & 0.416 & 0.006 & 0.004 \\
    \textbf{Average} & 0.596 & 0.461 & 0.007 & 0.005 \\
    \textbf{Maximum} & 0.81 & 0.603 & 0.015 & 0.010 \\
    \textbf{SD} & 0.05654 & 0.03962 & 0.00105 & 0.00076 \\
    \textbf{IQR} & 0.06212 & 0.04843 & 0.00092 & 0.00047 \\
    \bottomrule
  \end{tabular}
  \label{tab:validation:performance:reasoning:prediction:layers}
\end{table}

Fig. \ref{fig:validation:accuracy:reasoning:prediction:layers} plots the average accuracy and F1-score metrics for classification into the Security Categories of the \emph{ns} domain in Layer 1. It is shown per estimator and \textit{k-fold}, as obtained from \texttt{scikit} built-in score and classification report functions.
Table \ref{tab:validation:accuracy:reasoning:prediction:layers} gathers averaged classification metrics, such as the Area Under the Curve (AUC).
Both estimators provide an average 85\% accuracy.
Table \ref{tab:validation:accuracy:reasoning:prediction:categories} presents the classification accuracy per Security Category in the \emph{ns} Security Domain, per estimator and for three validation scenarios: (i) VS1, training and testing with expert-generated CPD data (whose extended classification metrics are in Table \ref{tab:validation:accuracy:reasoning:prediction:layers}); (ii) VS2, training with expert CPD and testing with partly random CPD values (i.e. randomised $r_c$); and (iii) VS3, both training and testing with partly random CPD values.
Classification (i) performs consistently worse for the \textit{AC} Security Category; (ii) loses accuracy (minimum of 78\%) in VS3; and (iii) tends to perform best with the Bayesian estimator (except for VS3, where MLE slightly outperforms due to \textit{NV} and \textit{SC} classification).

\begin{table}[t]
  \vspace{0.4em}
  \caption{Classification metrics for Layer 1, domain: ns}
  \centering
  \begin{tabular}{lccccc}
    \toprule
    \textbf{Estimator} & \textbf{Accuracy} & \textbf{AUC} & \textbf{Precision} & \textbf{Recall} & \textbf{F1-score} \\
    \hline \\[-0.5em]
    \textbf{Bayesian} & 0.85659 & 0.87844 & 0.88975 & 0.8856 & 0.88461 \\
    \textbf{MLE} & 0.85447 & 0.89224 & 0.85325 & 0.85447 & 0.85116 \\
    \bottomrule
  \end{tabular}
  \label{tab:validation:accuracy:reasoning:prediction:layers}
\end{table}

Finally, the accuracy and performance of the Gemini \textit{Flash Large} Language Model (LLM) were tested. The BN structure and requirements were passed as prompt.
Disparate results were obtained: 2.0 showed 0\% accuracy and performance time between 0.103 and 0.896 seconds for approximated inference results. 2.5 and 2.5 Pro models attained 100\% accuracy, with times up to 4-5 seconds due to its verbose multi-step reasoning process.
When compared to more focused, direct works such as the proposed one, the most accurate models are hindered by token limitation (2.5 Pro) and too long reasoning times.

\begin{table}[t]
  \caption{Accuracy per Category for Layer 1, domain: ns}
  \centering
    \begin{tabular}{lcccccc}
    \toprule
    \textbf{Estimator} & \multicolumn{3}{c}{\textbf{Bayesian}} & \multicolumn{3}{c}{\textbf{MLE}}  \\
    \textbf{Scenario}  & \textbf{VS1}       & \textbf{VS2}      & \textbf{VS3}      & \textbf{VS1}     & \textbf{VS2}     & \textbf{VS3}     \\
    \hline \\[-0.5em]
    \textbf{AC}        & 0.7516   & 0.7184  & 0.5574  & 0.6284 & 0.6002 & 0.5462 \\
    \textbf{NV}        & 0.9328   & 0.9326  & 0.835   & 0.952  & 0.9542 & 0.8496 \\
    \textbf{SC}        & 0.9724   & 0.9726  & 0.95    & 0.983  & 0.9812 & 0.9722 \\
    \textbf{\textit{(Average)}} & 0.8566   & 0.8745  & 0.7808  & 0.8545 & 0.8452 & 0.7893 \\
    \bottomrule
    \end{tabular}
  \label{tab:validation:accuracy:reasoning:prediction:categories}
\end{table}

%% file: sections/07-conclusions.tex
\section{Conclusion}
\label{sec:conclusions}

This work proposes an easy-to-use and extensible BN-based Security DSS framework. It gathers minimal end-user, text-based requirements from the CIA triad, constructs on-the-fly BNs from bundled model definitions and infers Security Mechanisms meeting such requirements.
A first set of results is presented, per reasoning pattern and processing.

Identified improvements on the current design are scaling to larger models and introducing a graphical interface to process the Security Requirements' weights through e.g. Kiviat diagrams or linked sliders.
On the other hand, evaluation can compare against other inference algorithms and statistical estimators; and the DSS can adopt Causal Influence Diagrams, MOO techniques with utility functions or Machine Learning (ML) ensemble learning approaches.

%% file: references.bib
@book{Lenschow_Kalia_Sandler_El-Assal_2024,
	title        = {Barclays’ 1H24 CIO Survey: 2024 Outlook Sustained},
	author       = {Lenschow, Raimo and Kalia, Saket and Sandler, Ross and El-Assal, Ramsey},
	year         = 2024,
	month        = {apr},
	url          = {https://a.storyblok.com/f/148396/x/f3dfd0d41a/barclays_cio_survey_2024.pdf},
	institution  = {Barclays},
	publisher    = {Barclays},
	language     = {en}
}

@article{LOUREIRO202113,
	title        = {Security misconfigurations and how to prevent them},
	author       = {Sergio Loureiro},
	year         = 2021,
	journal      = {Network Security},
	volume       = 2021,
	number       = 5,
	pages        = {13--16},
	doi          = {10.1016/S1353-4858(21)00053-2},
	issn         = {1353-4858}
}

@inproceedings{10.1145/3243734.3243794,
	title        = {Investigating System Operators' Perspective on Security Misconfigurations},
	author       = {Dietrich, Constanze and Krombholz, Katharina and Borgolte, Kevin and Fiebig, Tobias},
	year         = 2018,
	booktitle    = {Proceedings of the 2018 ACM SIGSAC Conference on Computer and Communications Security},
	location     = {Toronto, Canada},
	publisher    = {Association for Computing Machinery},
	address      = {New York, NY, USA},
	series       = {CCS '18},
	pages        = {1272–1289},
	doi          = {10.1145/3243734.3243794},
	isbn         = 9781450356930,
	numpages     = 18
}

@book{Jensen_Nielsen_2007,
	title        = {Bayesian networks and decision graphs},
	author       = {Jensen, Finn Verner and Nielsen, Thomas D.},
	year         = 2007,
	month        = {feb},
	publisher    = {Springer},
	address      = {New York},
	series       = {Information science and statistics},
	isbn         = {978-0-387-68281-5},
	edition      = {2nd ed},
	callnumber   = {519.542},
	collection   = {Information science and statistics},
	language     = {en}
}

@inproceedings{8842769,
	title        = {Decision Support System for Identification and Security Management of Essential and Digital Services},
	author       = {Kamola, Mariusz and Jaskóła, Przemysław and Amanowicz, Marek},
	year         = 2019,
	booktitle    = {2019 International Conference on Military Communications and Information Systems (ICMCIS)},
	volume       = {},
	number       = {},
	pages        = {1--7},
	doi          = {10.1109/ICMCIS.2019.8842769}
}

@article{Ahmed_Hossain_Fazio_Lezzi_Islam_2024,
	title        = {A decision support model for assessing and prioritization of industry 5.0 cybersecurity challenges},
	author       = {Ahmed, Ifaz and Hossain, Niamat Ullah Ibne and Fazio, Steven A and Lezzi, Marianna and Islam, Md. Saiful},
	year         = 2024,
	month        = {jan},
	journal      = {Sustainable Manufacturing and Service Economics},
	volume       = 3,
	pages        = 100018,
	doi          = {10.1016/j.smse.2024.100018},
	issn         = {2667-3444},
	abstractnote = {The world is adopting the Industry 5.0 paradigm to increase human centricity, sustainability, and resilience in efficient, optimized, and profitable manufacturing systems. With benefits, however, come increased risks of economic and physical loss, driving the need for continuous improvement of Industry 5.0 cybersecurity. Implementation and advancement of adequate cybersecurity have created challenges that have been identified in the literature. In this study, key Industry 5.0 cybersecurity challenges and related sub-challenges are highlighted based on a literature review. Graph Theory and Matrix Approach (GTMA) is employed to analyze the challenges and determine relative importance based on permanent values of the variable permanent matrix (VPM). The results identify the most important Industry 5.0 cybersecurity challenges and reveal Industry 5.0 firms should primarily concentrate on supply chain vulnerabilities to decrease data loss and hacking in the organization’s supply chain network. This study also recommends that executives and lawmakers acquire knowledge regarding cybersecurity challenges and prepare to deal with them. Addressing these and other subsequently prioritized challenges—the top five rounded out with emergent cybersecurity trends, non-availability of cybersecurity curriculum in education, embedded technical constraints, and absence of skilled employees and training—will lead the methodical development of holistic, robust cybersecurity programs. Firms accepting of this reality may implement such programs to mitigate evolving cyber-risk towards harnessing and sustaining the benefits of novel Industry 5.0 technologies.}
}

@inproceedings{10.1007/978-3-642-16644-0_39,
	title        = {Fusion of Bayesian and Ontology Approach Applied to Decision Support System for Critical Infrastructures Protection},
	author       = {Kozik, Rafa{\l} and Chora{\'{s}}, Micha{\l} and Ho{\l}ubowicz, Witold},
	year         = 2010,
	booktitle    = {Mobile Lightweight Wireless Systems},
	publisher    = {Springer Berlin Heidelberg},
	address      = {Berlin, Heidelberg},
	pages        = {451--463},
	isbn         = {978-3-642-16644-0},
	editor       = {Chatzimisios, Periklis and Verikoukis, Christos and Santamar{\'i}a, Ignacio and Laddomada, Massimiliano and Hoffmann, Oliver}
}

@article{Sawik_2022,
	title        = {A linear model for optimal cybersecurity investment in Industry 4.0 supply chains},
	author       = {Sawik, Tadeusz},
	year         = 2022,
	month        = {feb},
	journal      = {International Journal of Production Research},
	publisher    = {Taylor \& Francis},
	volume       = 60,
	number       = 4,
	pages        = {1368–1385},
	doi          = {10.1080/00207543.2020.1856442},
	issn         = {0020-7543},
	abstractnote = {This paper presents a mixed integer linear programming formulation for optimisation of cybersecurity investment in Industry 4.0 supply chains. Using a recursive linearisation procedure, a complex nonlinear stochastic combinatorial optimisation model with a classical exponential function of breach probability is transformed into its linear equivalent. The obtained linear optimisation model is capable of selecting optimal portfolio of security safeguards to minimise cybersecurity investment and expected cost of losses from security breaches in a supply chain. The new efficiency measures of cybersecurity investment are introduced: cybersecurity value and cybersecurity ratio. In addition, the proposed linear model has been enhanced for the Hurwicz-type, best–worst criterion to minimise a convex combination of the minimal and the maximal supply chain node vulnerability, under limited budget. The resulting compromise cybersecurity investment aims at balancing vulnerability over the entire supply chain, independent of cyberattack probabilities and potential losses by security breaches, thereby hardening the weaker critical nodes. The findings indicate a crucial role of intrinsic vulnerability, determined by the architecture of Industry 4.0 supply chain, and highlight ‘design for cybersecurity’ as an important emerging area of research.}
}

@article{OZDEMIRSONMEZ2022102865,
	title        = {Decision support for healthcare cyber security},
	author       = {Ferda {Özdemir Sönmez} and Chris Hankin and Pasquale Malacaria},
	year         = 2022,
	journal      = {Computers \& Security},
	volume       = 122,
	pages        = 102865,
	doi          = {10.1016/j.cose.2022.102865},
	issn         = {0167-4048}
}

@article{Simsek_Dag_Coussement_Kibis_Asilkalkan_Ragothaman_2025,
	title        = {A decision support framework for misstatement identification in financial reporting: A hybrid tree-augmented Bayesian belief approach},
	author       = {Simsek, Serhat and Dag, Ali and Coussement, Kristof and Kibis, Eyyub Y. and Asilkalkan, Abdullah and Ragothaman, Srinivasan},
	year         = 2025,
	month        = feb,
	journal      = {Decision Support Systems},
	volume       = 189,
	pages        = 114369,
	doi          = {10.1016/j.dss.2024.114369},
	issn         = {0167-9236},
	abstractnote = {Over a six-year period, employees and managers at Wells Fargo created 3.5 million false deposit and credit card accounts resulting in $4.8 billion in fines. Following this incident, there has been a newfound focus on effective internal controls. The purpose of the current study is to improve misstatement identification by formulating a novel hybrid decision support framework to a) accurately predict financial misstatements and frauds, b) build a parsimonious model by employing a comprehensive variable selection procedure without hurting (in contrast, potentially improving) the model’s prediction power, c) uncover the conditional inter-dependencies between the predictors via a Bayesian-belief based probabilistic network, and d) provide stakeholders with a firm-specific MWIC risk score. In an extensive real-life experimental setup, we validate our decision support system and find that the Tree-Augmented Bayesian Belief Network (TAN) model provides high misstatement identification accuracy results when the variables are selected through the Genetic Algorithm (GA) that employs Random Forests (RF) as the classification algorithm (AUC of 0.856 by employing only 5 out of 23 potential variables). Financial experts and stakeholders can use the probabilistic scores provided, while their intuition/incentive should collaborate with prediction models to make final decision on the cases where the model is not confident enough (i.e., when the probabilistic scores are close to 50/50). These insights enable stakeholders to improve the early warning systems for MWIC and financial misstatements and therefore potential frauds.}
}

@article{Wang_Neil_Fenton_2020,
	title        = {A Bayesian network approach for cybersecurity risk assessment implementing and extending the FAIR model},
	author       = {Wang, Jiali and Neil, Martin and Fenton, Norman},
	year         = 2020,
	month        = {feb},
	journal      = {Computers \& Security},
	volume       = 89,
	pages        = 101659,
	doi          = {10.1016/j.cose.2019.101659},
	issn         = {0167-4048},
	abstractnote = {Quantitative risk assessment can play a crucial role in effective decision making about cybersecurity strategies. The Factor Analysis of Information Risk (FAIR) is one of the most popular models for quantitative cybersecurity risk assessment. It provides a taxonomic framework to classify cybersecurity risk into a set of quantifiable risk factors and combines this with quantitative algorithms, in the form of a kind of Monte Carlo (MC) simulation combined with statistical approximation techniques, to estimate cybersecurity risk. However, the FAIR algorithms restrict both the type of statistical distributions that can be used and the expandability of the model structure. Moreover, the applied approximation techniques (including using cached data and interpolation methods) introduce inaccuracy into the FAIR model. To address restrictions of the FAIR model, we develop a more flexible alternative approach, which we call FAIR-BN, to implement the FAIR model using Bayesian Networks (BNs). To evaluate the performance of FAIR and FAIR-BN, we use a MC method (FAIR-MC) to implement calculations of the FAIR model without using any of the approximation techniques adopted by FAIR, thus avoiding the corresponding inaccuracy that can be introduced. We compare the empirical results generated by FAIR and FAIR-BN against a large number of samples generated using FAIR-MC. Both FAIR and FAIR-BN provide consistent results compared with FAIR-MC for general cases. However, the FAIR-BN achieves higher accuracy in several cases that cannot be accurately modelled by the FAIR model. Moreover, we demonstrate that FAIR-BN is more flexible and extensible by showing how it can incorporate process-oriented and game-theoretic methods. We call the resulting combined approach “Extended FAIR-BN” (EFBN) and show that it has the potential to provide an integrated solution for cybersecurity risk assessment and related decision making.}
}

@article{Hunte_Neil_Fenton_2022,
	title        = {A causal Bayesian network approach for consumer product safety and risk assessment},
	author       = {Hunte, Joshua L. and Neil, Martin and Fenton, Norman E.},
	year         = 2022,
	month        = {feb},
	journal      = {Journal of Safety Research},
	volume       = 80,
	pages        = {198–214},
	doi          = {10.1016/j.jsr.2021.12.003},
	issn         = {0022-4375},
	abstractnote = {Introduction: Product risk assessment is the overall process of determining whether a product is judged safe for consumers to use. Among several methods for product risk assessment, RAPEX is the primary one used by regulators in the UK and EU. Despite its widespread use we identify several limitations of RAPEX, including a limited approach to handling uncertainty, especially in the absence of data, and the inability to incorporate causal explanations for using and interpreting the data. Method: We develop a Bayesian Network (BN) model to provide an improved systematic method for product risk assessment that resolves the identified limitations with RAPEX. BNs are a rigorous, normative method for modelling uncertainty and causality which are already used for risk assessment in domains such as medicine and finance, as well as critical systems generally. Results: We use the BN approach to demonstrate risk assessments for products where relevant test and product instance data are and are not available. Whereas RAPEX can only produce results given relevant data, the BN approach produce results for products with and with no relevant data – replicating RAPEX in the former and providing deeper insights in both cases. Conclusion: The BN approach is powerful and flexible for systematic product risk assessment. While it can complement more traditional methods like RAPEX, it is able to provide quantified, auditable assessments in situations where such methods cannot because of lack of data. Practical Applications: Safety regulators, manufacturers, and risk professionals can use the BN approach for all types of consumer product risk assessment, including for novel products or products with little or no historical data. They can also use it to validate the results of existing methods when data becomes available. It informs risk management decisions and helps understand the effect of those decisions on the consumer risk perception.}
}

@article{Hunte_Neil_Fenton_2024,
	title        = {A hybrid Bayesian network for medical device risk assessment and management},
	author       = {Hunte, Joshua L. and Neil, Martin and Fenton, Norman E.},
	year         = 2024,
	month        = {jan},
	journal      = {Reliability Engineering \& System Safety},
	volume       = 241,
	pages        = 109630,
	doi          = {10.1016/j.ress.2023.109630},
	issn         = {0951-8320},
	abstractnote = {Risk analysis methods for medical devices, including fault tree analysis, have limitations such as handling uncertainty and providing reasonable risk estimates with limited or no testing data. To address these limitations, this paper proposes a novel systematic method for medical device risk management using hybrid Bayesian networks (BNs). We apply the method to a Defibrillator device to demonstrate the process involved for risk management during production and post-production using 4 different scenarios: (1) where there are available testing data; (2) where there are limited or no testing data; (3) where it is a completely new device with no testing data; (4) where we are reassessing the risk of a previous model on the market based on reported hazards and injuries. In each scenario, the BN model, for the available data, provides the full probability of failure per demand distribution for each category of injury severity (fatal, critical, major, minor, negligible) and the probabilities associated with various risk acceptability criteria. The model results are validated using publicly available data for the LIFEPAK 1000 Defibrillator (PN: 320371500XX), which was recalled by Physio-Control in 2017. The results show that the device would fail the acceptability criteria for probability of fatal injury.}
}

@article{doi:10.1177/1548512916683451,
	title        = {Markov Chain modeling of cyber threats},
	author       = {Ross Gore and Jose Padilla and Saikou Diallo},
	year         = 2017,
	journal      = {The Journal of Defense Modeling and Simulation},
	volume       = 14,
	number       = 3,
	pages        = {233--244},
	doi          = {10.1177/1548512916683451}
}

@article{Zadeh_Jeyaraj_2022,
	title        = {A multistate modeling approach for organizational cybersecurity exploration and exploitation},
	author       = {Zadeh, Amir and Jeyaraj, Anand},
	year         = 2022,
	month        = {nov},
	journal      = {Decision Support Systems},
	series       = {Business and Government Applications of Text Mining \& Natural Language Processing (NLP) for Societal Benefit},
	volume       = 162,
	pages        = 113849,
	doi          = {10.1016/j.dss.2022.113849},
	issn         = {0167-9236},
	abstractnote = {This study examines the dynamic stages of exploration and exploitation efforts by organizations in their cybersecurity responses using multistate modeling. Using textual data from the annual 10-K reports of S&P 100 organizations, this study uses a combination of text analytics and Markov chain approach to quantify exploration and exploitation in organizational cybersecurity responses. The study models two and four states of exploration and exploitation based on their cybersecurity responses over time and uses a continuous-time Markov chain approach to analyze transitions between states as organizations adapt their responses over time to achieve ambidexterity. The two-state Markov model focuses on the firm-level Exploration and Exploitation states whereas the four-state model captures deeper levels of exploration and exploitation by considering Surviving, Investigating, Reinforcing, and Balancing as possible states for exploration and exploitation of cybersecurity responses. We analyze the dynamics of organizational exploration-exploitation behaviors by modeling longitudinal transition probabilities across different states. Implications for research and practice are discussed.},
	collection   = {Business and Government Applications of Text Mining \& Natural Language Processing (NLP) for Societal Benefit}
}

@article{PAUL2021349,
	title        = {Decision support model for cybersecurity risk planning: A two-stage stochastic programming framework featuring firms, government, and attacker},
	author       = {Jomon A. Paul and Minjiao Zhang},
	year         = 2021,
	journal      = {European Journal of Operational Research},
	volume       = 291,
	number       = 1,
	pages        = {349--364},
	doi          = {10.1016/j.ejor.2020.09.013},
	issn         = {0377-2217}
}

@article{SCHMIDT2021107093,
	title        = {Risk management for cyber-infrastructure protection: A bi-objective integer programming approach},
	author       = {Adam Schmidt and Laura A. Albert and Kaiyue Zheng},
	year         = 2021,
	journal      = {Reliability Engineering \& System Safety},
	volume       = 205,
	pages        = 107093,
	doi          = {10.1016/j.ress.2020.107093},
	issn         = {0951-8320}
}

@book{Koller_Friedman_2010,
	title        = {Probabilistic graphical models: principles and techniques},
	author       = {Koller, Daphne and Friedman, Nir},
	year         = 2010,
	publisher    = {MIT Press},
	address      = {Cambridge, Mass.},
	series       = {Adaptive computation and machine learning},
	isbn         = {978-0-262-01319-2},
	url          = {http://mcb111.org/w06/KollerFriedman.pdf},
	edition      = {Nachdr.},
	collection   = {Adaptive computation and machine learning},
	language     = {en}
}

@phdthesis{Requejo_Castro_Perez_2021,
	title        = {Data–driven Bayesian networks modelling to support decision–making: application to the context of Sustainable Development Goal 6 on water and sanitation},
	author       = {Requejo Castro, David and Pérez Foguet, Agustí and Giné Garriga, Ricard},
	year         = 2021,
	month        = {jul},
	doi          = {10.5821/dissertation-2117-351115},
	rights       = {http://creativecommons.org/licenses/by-nc-sa/4.0/},
	abstractnote = {We live in a complexand interconnected world which permeates ditterent scales. sectors or decision problems. This fact is acknowledged by the United Nations 2030 Agenda for Sustainable Development, which underscores current global challenges, recognizes their interconnectivity and calls for international action. lt is recognized that the connected nature of the issues we currently face have been tackled by “silo” approaches, separating the complexities ofthe real-world into specialized disciplines. fields of research, institutions and ministries, each one focused on a fraction of the overall truth. Similarly, it is widely recognized the need of a major shift in decision-making processes towards more holistic and integrated approaches. Evidence-based decrsion-making involves complexprocesses ofconsidering a wide range of information of different nature. Nowadays, available data can support these processes, but methodologies to effectively integrate these data are lacking. With the aim to contribute in this direction, this thesis focuses on the increasing use of Bayesian Networks (BNs) modelling as an approach to accom m odate com plex problem s and to support decis ion-making. Com mon practica em ploys separately expert knowledge and empirical data to build and apply associated models. Des pite of the demonstrated utility of this practica, in an era where the data are bigger, faster and more detailed than even before, there is still room for further exploration. Thus, this dissertation proposes a data-driven Bayesian Networks approach to combine expert opinion and quantitative data to support informad decision-making. We propose two systematic methods to this end. First. we use our approach to replicate composite indicators (Cl)-based conceptual frameworks, which represent expert knowledge. through the use of structure learning algorithms, which characterizes this data-driven Bayesian Networks approach. Second, we use our approach to identify interlinkages associated with a complex context, coupled with a statistical technique (i.e. bootstrapping) to reduce results uncertainty and with a comprehensive result robustness analysis (i.e. expert knowledge). For testing and validating the proposed approach, this thesis takes the Sustainable Development Goal 6 embedded on the 2030 Agenda as a reference point, with particular attention to the water, sanitation and hygier:ie sector. Our results emphasize the likely utility of the data-driven Bayesian Networks approach adopted. First. it allows the integration of both expert knowledge and data availability when dealing with BNs modelling, and it accurately replicates (Cl)-based conceptual frameworks. As added values, this combination improves model inference capacity, it reduces and quantifies the key variables that explµin a pre-defined objective variable (implying important advantages in data updating), and it identifies the interlinkages among the variables considerad (which might enhance more integrated actions). Second, the approach adopted is useful to accommodate a thorough analysis and interpretation of the complexities and interdependencias of any context at hand. As added values, interlinkages identification is spurred on by the available data and this identification makes the approach more suitable than the use of composite indicators. Third, the systematic nature of the methodological contributions associated with the proposed approach can be adapted to different complexproblems. Thus, it might expand and deepen the knowledge about the validity, reliability and accuracy of using BNs modelling. Vivimos en un mundo complejo e interconectado que impregna diferentes escalas. sectores o problemas de decisi ón. Esta visión es destacada por la Agenda 2030 para el Desarrollo Sostenible de las Naciones Unidas, que además pone de manifiesto los desafíos globales actuales. reconoce su interconexión y hace una llamada a la acción internacional. Por otro lado. es ampliamente reconocido que la naturaleza conectada de los problemas a los que nos enfrentamos actualmente se ha abordado mediante enfoques “estancos”. separando las complejidades del mundo real en disciplinas especializadas. campos de investigación, instituciones y ministerios. cada uno centrado en una parte de la verdad. De igual manera. es reconocida la necesidad de un cambio de paradigma en los procesos de toma de decisiones hacia enfoques m ás holísticos e integrados. La toma de decisiones basada en evidencias lleva implícita procesos complejos en los que se integran una amplia gama de información de diferente naturaleza. Hoy en día, los datos cuantitativos disponibles pueden respaldar estos procesos. pero faltan metodologías para integrar estos datos de manera efectiva. Con el objetivo de contribuir en esta dirección, esta tesis se centra en el uso de modelos de Redes Bayesianas (BNs), como un enfoque válido para abordar problemas complejos y, en última instancia, para apoyar la toma de decisiones. En la práctica, se emplea comúnmente por separado el conocimiento de expertos y los datos empíricos para construir y aplicar estos modelos. A pesar de la utilidad de esta práctica, en una era en la que los datos son má¿ numerosos, más rápidos y más detallados que antes, hay espacio para explorar hasta dónde pueden llegar estos datos. En este sentido, esta tesis propone un enfoque de Redes Bayesianas basadas en los datos que permite combinar el conocimiento experto y la información cuantitativa existente para, en última instancia, apoyar la toma de decisiones. Se proponen dos métodos sistemáticos para tal fin. En primer lugar, se emplea dicho enfoque para replicar marcos conceptuales basados en indicadores compuestos (IC), que representan el conocimiento experto, mediante el uso de algoritmos de aprendizaje de estructuras, que caracteriza este enfoque de Redes Bayesianas basado en datos. En segundo lugar, se utiliza el enfoque propuesto para identificar las interrelaciones existentes dentro de un contexto complejo, junto con una técnica estadística (bootstrapping) para reducir la incertidumbre de los resultados y con un análisis bibliográfico exhaustivo (conocimiento experto) para demostrar la robustez de los resultados obtenidos. Para testear y validar el enfoque propuesto, esta tesis toma como punto de referencia el Objetivo de Desarrollo Sostenible 6 que forma parte de la Agenda 2030, con especial atención al sector del agua, saneamiento e higiene. Nuestros resultados ponen de manifiesto la potencial utilidad del enfoque adoptado. Primero, este enfoque permite la integración de conocimiento experto y de información cuantitativa a la hora de construir las RBs, y replica con precisión los marcos conceptuales basados en IC. Como valor añadido, esta combinación mejora la capacidad de inferencia del modelo, y reduce y cuantifica las variables clave que explican una variable objetivo predefinida. En segundo lugar, el enfoque adoptado es útil para dar cabida a un análisis e interpretación exhaustivos de las complejidades e interdependencias de cualquier contexto en cuestión. Como valor añadido, la identificación de las interconexiones se realiza exclusivamente en base a los datos disponibles. Considerar dichas interconexiones hace que este enfoque sea más adecuado que el uso de IC. En tercer lugar. la naturaleza sistemática dé las contribuciones metodológicas asociadas al enfoque propuesto puede adaptarse a diferentes problemas complejos. En este sentido, se considera que se contribuye a ampliar el conocimiento sobre la validez, fiabilidad y precisión del uso de modelos de Redes Bayesianas.},
	school       = {Universitat Politècnica de Catalunya},
	language     = {en}
}

@inbook{L_Crowley_Reasoning_BN_2018,
	title        = {Reasoning with Bayesian Networks},
	author       = {L. Crowley, James},
	year         = 2018,
	month        = {apr},
	url          = {http://crowley-coutaz.fr//jlc/Courses/2017/ENSI2.SIRR/ENSI2.SIRR.S18.pdf}
}

@misc{Reasoning_with_Bayesian_Belief_Networks_2016,
	title        = {Reasoning with Belief Bayesian Networks},
	year         = 2016,
	month        = {apr},
	address      = {University of Maryland},
	url          = {https://courses.cs.umbc.edu/471/spring16/01/notes/15/bbn.pdf}
}

@inproceedings{Ankan2015,
	title        = {pgmpy: {P}robabilistic {G}raphical {M}odels using {P}ython},
	author       = {{A}nkur {A}nkan and {A}binash {P}anda},
	year         = {2015},
	booktitle    = {{P}roceedings of the 14th {P}ython in {S}cience {C}onference},
	pages        = {6--11},
	doi          = {10.25080/Majora-7b98e3ed-001},
	editor       = {{K}athryn {H}uff and {J}ames {B}ergstra}
}
